# Mutual Orbits and Masses of Six Transneptunian Binaries

W.M. Grundy[1], K.S. Noll[2], M.W. Buie[3], S.D. Benecchi[2],
D.C. Stephens[4], and H.F. Levison[5].

1. Lowell Observatory, 1400 W. Mars Hill Rd., Flagstaff AZ 86001.
2. Space Telescope Science Institute, 3700 San Martin Dr., Baltimore MD 21218.
3. Southwest Research Institute, 1050 Walnut St. #300, Boulder CO 80302; formerly at Lowell Observatory, 1400 W. Mars Hill Rd., Flagstaff AZ 86001.
4. Dept. of Physics and Astronomy, Brigham Young University, N283 ESC Provo UT 84602.
5. Southwest Research Institute, 1050 Walnut St. #300, Boulder CO 80302.

**ABSTRACT**

We present Hubble Space Telescope observations of six binary transneptunian systems: 2000 QL$_{251}$, 2003 TJ$_{58}$, 2001 XR$_{254}$, 1999 OJ$_4$, (134860) 2000 OJ$_{67}$, and 2004 PB$_{108}$.  The mutual orbits of these systems are found to have periods ranging from 22 to 137 days, semimajor axes ranging from 2360 to 10500 km, and eccentricities ranging from 0.09 to 0.55.  These orbital parameters enable estimation of system masses ranging from 0.2 to 9.7 × 10$^{18}$ kg.  For reasonable assumptions of bulk density (0.5 to 2.0 g cm$^{-3}$), the masses can be combined with visible photometry to constrain sizes and albedos.  The resulting albedos are consistent with an emerging picture of the dynamically "Cold" Classical sub-population having relatively high albedos, compared with comparably-sized objects on more dynamically excited orbits.

## 1. Introduction

As with binaries in other astrophysical settings, transneptunian binaries (TNBs) offer a way of measuring fundamental, but otherwise unobtainable physical properties such as masses and densities (e.g., Noll et al. 2008a). TNBs also offer potential insights into their dynamical environments over past history. They are relatively fragile systems which can be disrupted or have their mutual orbits altered by various external influences (e.g., Petit and Mousis 2004; Kern and Elliot 2006). To capitalize on the opportunities offered by TNBs will require knowledge of a large sample of their mutual orbits, enabling statistical comparisons among various sub-groupings. Toward that end, this paper describes Hubble Space Telescope (HST) observations of six TNB systems as listed in Table 1, leading to determination of the periods, semimajor axes, and eccentricities of their mutual orbits.

**Table 1.** Heliocentric orbital characteristics of the six TNB systems.

| TNB system number & designation | Mean heliocentric orbital elements[a] | | | Dynamical class[b] |
|---|---|---|---|---|
| | $a_\odot$ (AU) | $e_\odot$ | $i_\odot$ (°) | |
| 2000 QL$_{251}$ | 47.8 | 0.208 | 5.83 | Resonant 2:1 |
| 2003 TJ$_{58}$ | 44.5 | 0.094 | 1.31 | Classical |
| 2001 XR$_{254}$ | 43.0 | 0.024 | 2.66 | Classical |
| 1999 OJ$_4$ | 38.1 | 0.018 | 2.61 | Classical |
| (134860) 2000 OJ$_{67}$ | 42.9 | 0.014 | 1.33 | Classical |
| 2004 PB$_{108}$ | 45.1 | 0.107 | 19.19 | Scattered |

Table notes:

[a.] Averaged over a 10 Myr integration, with $i_\odot$ relative to the invariable plane, as described by Elliot et al. (2005).

[b.] Classifications are according to the Deep Ecliptic Survey system (DES; Elliot et al. 2005). The Gladman et al. (2008) system would identify 2004 PB$_{108}$ as a Classical object.

## 2. Data Acquisition, Reduction, and Orbit Determination

Data used in this paper were acquired by HST programs and 9386, 10514, 10800, and 11178, extending over Cycles 11 through 16. Various instruments, filters, and observing strategies were employed by these programs, but crucially, they all took multiple sequential images, dithered to improve sampling of the point spread function (PSF) of the telescope, to allow identification and exclusion of bad pixels and cosmic ray strikes, and to enable direct calculation of the precision of the photometry and relative astrometry derived from the data. Example single-frame images are shown in Fig. 1.

Two instruments were involved in the discovery of the binary nature and initial epoch astrometry of the TNBs in this paper. The Near Infrared Camera and Multi-Object Spectrometer (NICMOS; see Barker at al. 2007) was used by Noll et al. Cycle 11 program 9386 in a near-infrared color survey of transneptunian objects (TNOs) using the *F110W* and *F160W* filters[1] of NICMOS's NIC2 camera. That program serendipitously discovered a number of binary companions, providing single epoch relative astrometry at a relatively coarse pixel scale (Stephens and Noll 2006). The Advanced Camera for Surveys (ACS; Ford et al. 1996) was used in a pair of searches for TNO binary companions by Noll et al. in Cycles 14 and 15 (programs 10514 and 10800; see Noll et al. 2008b for details). To maximize sensitivity to faint and close-in companions, those programs used the high-throughput *CLEAR* filter combination of ACS's High Resolution Camera (HRC). The capabilities of ACS/HRC for relative astrometry of point sources separated by a small fraction of an arcsec benefit from the unusually fine pixel scale and an especially thorough calibration effort (e.g., Sirianni et al. 2005; Pavlovsky et al. 2006; Boffi et al. 2007).

---

[1] Broadband HST filter names mentioned in this paper (other than the self-explanatory *CLEAR* filter) take the

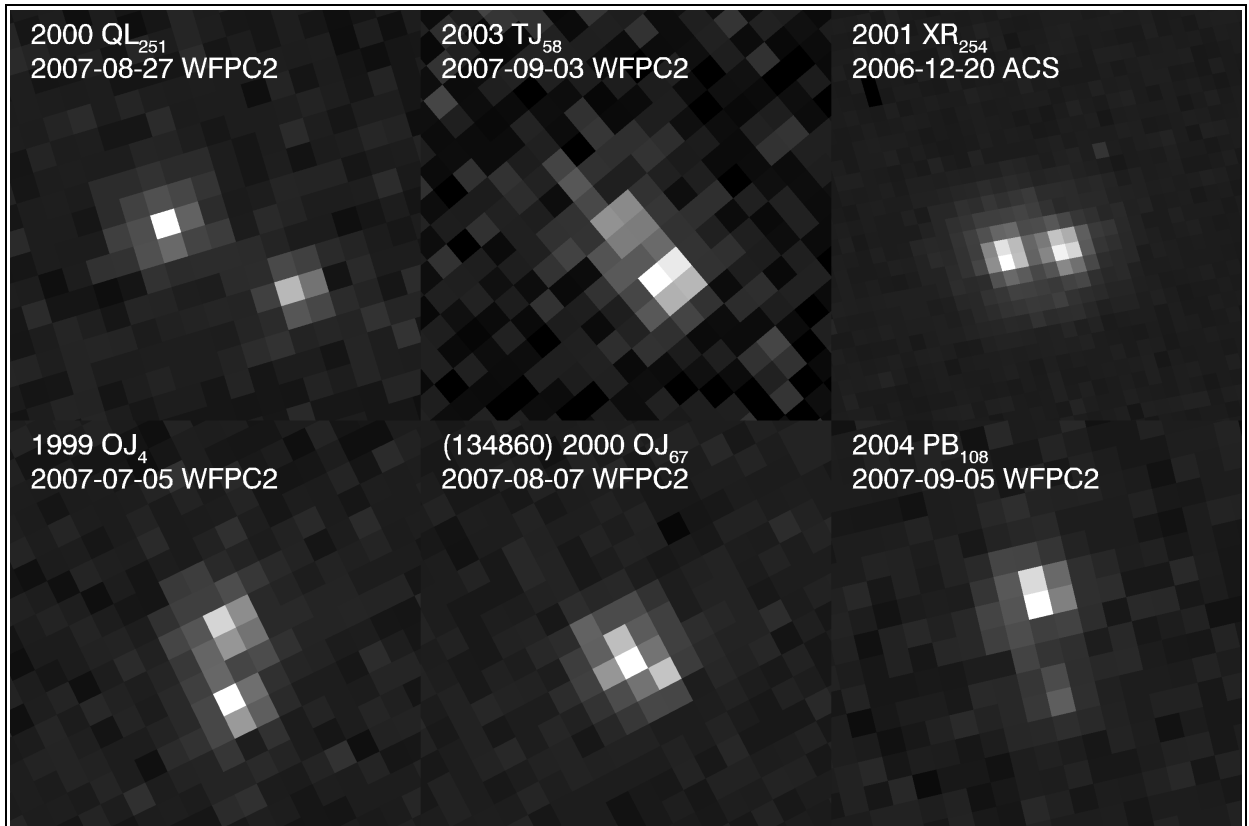

**Fig. 1:** Examples of single-frame HST images of our six TNB targets, stretched linearly and projected to sky-plane geometry with North up and East to the left. For scale, the width of this entire montage is 2 arcsec. Dithering is crucial with WFPC2 to fully sample the PSF, so for each visit we record images using four separate pointings per filter. Note the smaller, more distorted shapes of the ACS/HRC pixels when projected to the sky plane (upper right sub-panel). These smaller pixels do a better job of sampling the PSF.

Follow-up observations were obtained by Grundy et al. Cycle 16 program 11178. The untimely failure of ACS forced this program to use the older Planetary Camera of the Wide Field and Planetary Camera 2 (WFPC2/PC; McMaster et al. 2008). WFPC2/PC has no clear filter, so most observations were done using *F606W*, the most sensitive available filter (additional observations using the *F814W* filter to obtain colors are the subject of a companion paper, Benecchi et

---

form of "*F*", followed by the first 3 digits of the central wavelength in nanometers, followed by "*W*" for wide. *F606W*, *F814W*, *F110W*, and *F160W* filters have central wavelengths of 606, 814, 1100, and 1600 nm, respectively. Detailed filter response information can be found at http://www.stsci.edu/hst/HST_overview/instruments.

al. 2009, which describes in detail the extraction of photometry from WFPC2 images). Observing overheads with WFPC2/PC are considerably higher than with ACS, the pixel scale is coarser and does not sample the telescope's PSF as well as ACS/HRC does, and charge transfer efficiencies are inferior. These factors combine to reduce the signal precision and number of frames which can be obtained in a single HST orbit and to reduce the precision of the relative astrometry which can be extracted from them.

Our data reduction procedures rely on the detailed knowledge of HST's exceptionally stable PSF expressed in the Tiny Tim software package (e.g., Krist and Hook 2004). For each image frame, we simultaneously fit a pair of PSFs representing the primary and secondary body. The scatter of the measurements from the multiple dithered frames obtained during a single visit is used to estimate uncertainties on the mean for that visit. Details of our processing pipelines for NICMOS and for ACS/HRC data are published elsewhere (Stephens and Noll 2006; Grundy et al. 2008; Benecchi et al. 2009). For WFPC2/PC data, the procedure is comparable. Fitting is done in the data frame, minimizing $\chi^2$ for a $21 \times 21$ pixel region extracted from the full $800 \times 800$ data frame (see Fig. 2; smaller boxes are occasionally used for tighter pairs to avoid background clutter). The downhill simplex "amoeba" algorithm (Nelder and Mead 1965; Press et al. 1992) is used to simultaneously adjust the parameters of a synthetic image to minimize $\chi^2$ with respect to the data image. The synthetic image is produced from a high resolution Tiny Tim PSF generated for a solar spectral distribution and the appropriate region of the WFPC2/PC array. This PSF is re-scaled and re-sampled to simulate point sources at appropriate fluxes and locations. Real images often appear slightly blurred in comparison with synthetic PSFs. A variety of factors may contribute to this image degradation, including telescope focus variations ("breathing"), redder than Solar object colors, finite sizes of the observed objects, tracking errors, and perhaps other effects not yet recognized. To approximately account for these effects without adding numerous additional free parameters, we convolve the synthetic image with a 2-dimensional, rotationally-symmetric Gaussian smearing kernel, with a full width at half maximum generally smaller than a pixel. Note that this smearing Gaussian is distinct from the pixel-response function which we also include in generating the synthetic image. The sky background is fit separately, and for higher flux objects, the width of the smearing Gaussian is also fit separately, iterating between fitting the background level, the width of the smearing Gaussian, and the PSF locations and amplitudes until nothing changes. Fitting these components separately was found to improve numerical stability. These procedures usually result in goodness of fit to the image around unity, as measured by $\chi_\nu^2$ ($\chi^2$ reduced by the number of degrees of freedom). However, for other TNBs much brighter than the six subjects of this paper, $\chi_\nu^2$ tends to be greater than one. Final fitted pixel positions are converted to relative astrometry by means of the ORIENTAT header keyword and the *F606W* image scale of 45.5543 ± 0.0050 mas pixel$^{-1}$, (McMaster et al. 2008; for *F814W*, the value is 45.5743 ± 0.0050 mas pixel$^{-1}$). We identify the brighter object as the "primary" and the fainter one as the "secondary". This identification occasionally needs to be reversed after we learn more about the system, but that is a simple matter of a sign change. Uncertainties in the visit mean astrometry are estimated from the scatter of the measurements, imposing a 1 mas floor to avoid over-weighting visits which could happen to have small measurement scatters by chance.

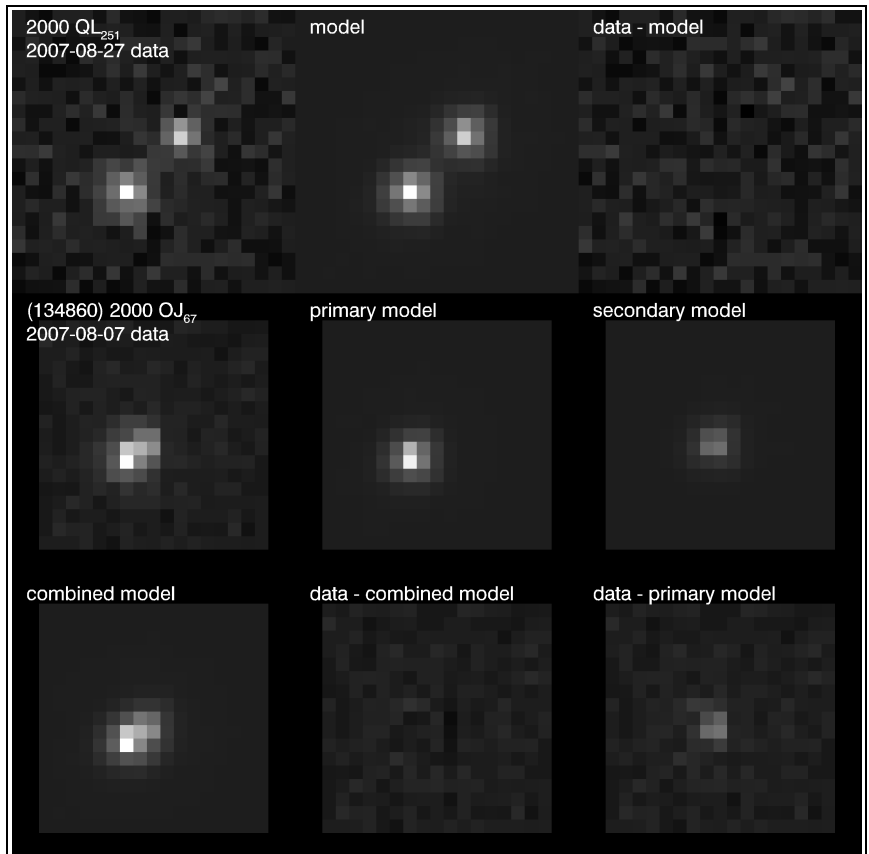

**Fig. 2:** Example single frame PSF fits for two of the WFPC2 images shown in Fig. 1. From left to right across the top row are the $21 \times 21$ pixel region of the data frame used in the fitting, the two-PSF model, and the residuals (data minus model, with average background added back in), all scaled the same, linearly. In the bottom two rows we used smaller $17 \times 17$ pixel sub-images because the separation between the two bodies was smaller. This example shows the primary and secondary components of the model image separately, and combined, along with residuals after subtracting the combined model and after subtracting only the primary component, leaving an image of the fainter secondary. Goodness of fit for these models are indicated by $\chi_\nu^2 = 0.8$ for the model in the top row and $\chi_\nu^2 = 0.9$ for the model in the bottom two rows.

For each TNB system discussed in this paper, from four to six follow-up visits were needed to determine the orbital period. Mean astrometric and photometric measurements and estimated 1-$\sigma$ uncertainties are listed in Table 2.

**Table 2.** Observational circumstances, relative astrometry, and photometry.

| Object and mean UT observation date[a] | HST instrument | $r$[b] (AU) | $\Delta$[b] (°) | $g$[b] | $\Delta x$[c] (arcsec) | $\Delta y$[c] | $V_{\text{primary}}$[d] (mag) | $V_{\text{secondary}}$[d] |
|---|---|---|---|---|---|---|---|---|
| **2000 QL$_{251}$** | | | | | | | | |
| 2006/07/25 10$^h$.7690 | ACS/HRC | 38.821 | 38.281 | 1.28 | −0.24089(045) | −0.10082(050) | - | - |
| 2007/07/15 22$^h$.7517 | WFPC2/PC | 38.943 | 38.576 | 1.40 | −0.10209(286) | −0.16995(124) | 23.950(22) | 24.243(51) |
| 2007/07/19 9$^h$.8100 | WFPC2/PC | 38.944 | 38.523 | 1.37 | −0.05006(118) | −0.15815(219) | 23.885(10) | 24.077(14) |
| 2007/08/05 11$^h$.7683 | WFPC2/PC | 38.950 | 38.287 | 1.14 | −0.02229(264) | +0.08140(189) | 24.030(05) | 23.823(32) |
| 2007/08/27 17$^h$.9405 | WFPC2/PC | 38.958 | 38.064 | 0.70 | −0.23335(194) | −0.12592(176) | 23.708(47) | 23.961(48) |
| 2008/08/25 8$^h$.6061 | WFPC2/PC | 39.088 | 38.221 | 0.77 | +0.07422(133) | −0.06801(148) | 23.785(21) | 23.913(20) |
| **2003 TJ$_{58}$** | | | | | | | | |
| 2006/11/22 5$^h$.8297 | ACS/HRC | 40.913 | 39.966 | 0.40 | +0.10010(122) | +0.06001(092) | - | - |
| 2007/08/12 1$^h$.6364 | WFPC2/PC | 40.939 | 41.427 | 1.24 | +0.09071(305) | −0.00587(239) | 24.335(39) | 24.780(57) |
| 2007/09/03 2$^h$.4072 | WFPC2/PC | 40.942 | 41.077 | 1.40 | +0.08640(182) | +0.10062(147) | 24.327(46) | 24.844(47) |
| 2007/10/22 16$^h$.8825 | WFPC2/PC | 40.947 | 40.295 | 1.06 | −0.05749(143) | +0.16160(227) | 24.205(25) | 24.722(49) |
| 2007/11/20 6$^h$.2408 | WFPC2/PC | 40.950 | 40.023 | 0.48 | −0.10744(215) | +0.07768(254) | 24.125(20) | 24.663(49) |

| Object and mean UT observation date[a] | HST instrument | $r$[b] (AU) | $\Delta$[b] (AU) | $g$[b] (°) | $\Delta x$[c] (arcsec) | $\Delta y$[c] (arcsec) | $V_{primary}$[d] (mag) | $V_{secondary}$[d] (mag) |
|---|---|---|---|---|---|---|---|---|
| **2001 XR$_{254}$** | | | | | | | | |
| 2006/12/20  6$^h$.6769 | ACS/HRC | 44.184 | 43.254 | 0.42 | −0.10693(098) | +0.01285(052) | - | - |
| 2007/09/17 11$^h$.1853 | WFPC2/PC | 44.174 | 44.588 | 1.18 | −0.30433(135) | +0.15170(107) | 23.305(52) | 23.522(52) |
| 2007/09/18  2$^h$.4936 | WFPC2/PC | 44.174 | 44.578 | 1.19 | −0.30475(194) | +0.15598(100) | 23.114(12) | 23.763(40) |
| 2007/09/21  7$^h$.6645 | WFPC2/PC | 44.174 | 44.527 | 1.21 | −0.31466(155) | +0.16542(100) | 23.282(52) | 23.720(35) |
| 2007/10/09  2$^h$.6269 | WFPC2/PC | 44.173 | 44.229 | 1.29 | −0.30501(140) | +0.19840(156) | 23.292(67) | 23.600(52) |
| 2007/12/05  9$^h$.9554 | WFPC2/PC | 44.171 | 43.371 | 0.75 | +0.10316(100) | +0.01377(188) | 23.440(63) | 23.378(44) |
| 2007/12/28 14$^h$.6145 | WFPC2/PC | 44.170 | 43.208 | 0.27 | −0.0178(180) | −0.02445(703) | 22.892(68) | 23.70(13) |
| **1999 OJ$_4$** | | | | | | | | |
| 2002/10/04 14$^h$.3591 | NICMOS | 38.174 | 37.559 | 1.19 | −0.03528(416) | +0.0706(101) | - | - |
| 2005/07/24 16$^h$.9878 | ACS/HRC | 38.103 | 37.164 | 0.59 | +0.09392(056) | +0.03652(057) | - | - |
| 2007/07/05 23$^h$.3058 | WFPC2/PC | 38.057 | 37.326 | 1.07 | −0.01522(222) | −0.14436(127) | 23.950(36) | 24.065(15) |
| 2007/07/07 10$^h$.5933 | WFPC2/PC | 38.057 | 37.309 | 1.05 | −0.02094(165) | −0.14481(100) | 23.915(19) | 24.035(18) |
| 2007/07/12 21$^h$.1017 | WFPC2/PC | 38.056 | 37.249 | 0.94 | −0.05148(281) | −0.13556(100) | 24.038(15) | 24.095(24) |
| 2007/08/01 22$^h$.2017 | WFPC2/PC | 38.055 | 37.091 | 0.48 | −0.07347(233) | +0.00338(446) | 23.903(13) | 23.967(39) |
| 2007/08/24  9$^h$.4206 | WFPC2/PC | 38.054 | 37.045 | 0.10 | +0.10699(247) | +0.00925(137) | 23.875(39) | 24.02(15) |
| **(134860) 2000 OJ$_{67}$** | | | | | | | | |
| 2003/06/25 17$^h$.5977 | NICMOS | 42.632 | 41.993 | 1.07 | −0.00044(283) | −0.08217(133) | - | - |
| 2007/07/16  7$^h$.6350 | WFPC2/PC | 42.718 | 41.902 | 0.82 | −0.06774(209) | −0.02806(174) | 23.340(30) | 24.167(57) |
| 2007/07/19 19$^h$.2850 | WFPC2/PC | 42.719 | 41.869 | 0.75 | −0.05977(100) | +0.03860(225) | 23.425(18) | 23.805(19) |
| 2007/08/06  6$^h$.1017 | WFPC2/PC | 42.720 | 41.747 | 0.39 | −0.05660(104) | −0.04953(117) | 23.285(24) | 23.943(18) |
| 2007/08/07  7$^h$.9864 | WFPC2/PC | 42.720 | 41.742 | 0.37 | −0.06764(333) | −0.02881(371) | 23.214(19) | 24.014(40) |
| 2008/05/04 22$^h$.1186 | WFPC2/PC | 42.736 | 43.023 | 1.29 | +0.00071(127) | +0.07146(337) | 23.627(64) | 23.73(11) |
| **2004 PB$_{108}$** | | | | | | | | |
| 2006/08/04  8$^h$.5017 | ACS/HRC | 43.500 | 42.639 | 0.71 | −0.15686(249) | +0.08128(134) | - | - |
| 2007/07/11 13$^h$.1350 | WFPC2/PC | 43.414 | 42.845 | 1.12 | +0.25528(156) | +0.26392(565) | 23.858(29) | 25.284(60) |
| 2007/07/15 13$^h$.4950 | WFPC2/PC | 43.412 | 42.789 | 1.07 | +0.22867(187) | +0.30690(188) | 23.913(32) | 25.307(82) |
| 2007/07/21 17$^h$.6767 | WFPC2/PC | 43.411 | 42.709 | 0.98 | +0.17984(181) | +0.36205(164) | 23.845(19) | 25.039(46) |
| 2007/09/05 20$^h$.3933 | WFPC2/PC | 43.399 | 42.397 | 0.15 | −0.02921(204) | −0.18204(313) | 23.706(23) | 25.095(32) |
| 2007/10/02  7$^h$.1825 | WFPC2/PC | 43.393 | 42.488 | 0.57 | +0.29506(360) | +0.05212(376) | 23.875(30) | 24.985(40) |

Table notes:

[a.] Dates are average UT mid-times of observations at the location of the observer (HST).

[b.] The distance from the Sun to the target is $r$ and from the observer to the target is $\Delta$. The phase angle, the angular separation between the observer and Sun as seen from the target, is $g$.

[c.] Relative right ascension $\Delta x$ and relative declination $\Delta y$ are computed as $\Delta x = (\alpha_2 - \alpha_1)\cos(\delta_1)$ and $\Delta y = \delta_2 - \delta_1$, where $\alpha$ is right ascension, $\delta$ is declination, and subscripts 1 and 2 refer to primary and secondary, respectively. 1-$\sigma$ uncertainties in the final 3 digits are indicated in parentheses. These are estimated from the scatter among each set of dithered images with a floor of 1 mas for WFPC2 observations, as described in the text. The body we identify as the "primary" of the 2001 XR$_{254}$ system was identified as the "secondary" by Benecchi et al. (2009). Due to lightcurve variations, it happened to be the fainter of the two

at the time of the 2007/12/05 two-filter visit analyzed in that paper.

d. Mean *V*-band photometry of primary and secondary bodies during each visit were derived from WFPC2/PC *F606W* images as described by Benecchi et al. (2009). Visits lacking *V* photometry in Table 2 involved other instruments and filters or coincided with times when the two bodies were especially tightly blended. 1-$\sigma$ uncertainties of the mean in the final 2 digits are indicated in parentheses.

To minimize the number of follow-up observations required, we used a scheduling strategy inspired by the Monte Carlo Statistical Ranging approach of Virtanen et al. (2008, and references therein). Details are in a companion paper (Grundy et al. 2008). Briefly, random orbits consistent with existing observations were collected and used to explore what regions of orbital element space were consistent with the data. These Monte Carlo orbit collections were also used to determine what future observation times would be best for excluding the remaining possible orbital solutions, thereby homing in on the true one. Finally, the orbit collections were used to evaluate whether or not we had arrived at a unique solution, or still needed additional observations. For each TNB in this paper, only two small clumps of orbital element space are consistent with the data. These two clumps correspond to two orbital solutions which are mirror images of one another through the average sky plane, with essentially the same period, semimajor axis, and eccentricity but different orientations.

Using orbits from the Monte Carlo clouds of orbits as starting vectors, we used the "amoeba" algorithm to fit Keplerian orbits to the observed relative astrometry, accounting for time-variable geometry and light time delays between the observer and the TNB system. The elements we used were period $P$, semimajor axis $a$, eccentricity $e$, inclination $i$, mean longitude $\epsilon$ at a reference epoch, longitude of the ascending node $\Omega$, and longitude of periapsis $\varpi$, referenced to the J2000 equatorial frame. Uncertainties on fitted elements were derived by generating a thousand sets of randomized astrometric measurements consistent with the actual measurements and their uncertainties. We fitted an orbit to each of these sets, then used the scatter of the ensemble of fitted elements to estimate uncertainties. Grundy et al. (2007, 2008) and Noll et al. (2008) provide more details on our methods of orbit fitting and assessment of parameter uncertainties. We ignored potential effects of center-of-mass/center-of-light discrepancies, internal dissipation, non-spherical mass distributions, and external perturbations such as Solar tides.

Orbital solutions along with quantities derived from them are listed in Tables 3-8. Sky plane astrometry and projections of the orbits are shown in Fig. 3. Sky-plane residuals for relative astrometry from ACS/HRC and WFPC2/PC images are typically in the one to two milliarcsec range, as would be expected from our estimated astrometric uncertainties.

**Table 3.** Orbital parameters and 1-$\sigma$ uncertainties for 2000 QL$_{251}$.

| Parameter | | Orbit 1 ($\chi^2 = 0.37$) | Orbit 2 ($\chi^2 = 1.9$) |
|---|---|---|---|
| Fitted elements:[a] | | | |
| Period (days) | $P$ | 56.459 ± 0.018 | 56.443 ± 0.017 |
| Semimajor axis (km) | $a$ | 4991 ± 17 | 5014 ± 16 |
| Eccentricity | $e$ | 0.4874 ± 0.0062 | 0.4867 ± 0.0060 |
| Inclination[b] (deg) | $i$ | 127.78 ± 0.62 | 45.62 ± 0.66 |

| Parameter | | Orbit 1 ($\chi^2 = 0.37$) | Orbit 2 ($\chi^2 = 1.9$) |
|---|---|---|---|
| Mean longitude[b] at epoch[c] (deg) | $\epsilon$ | 146.1 ± 1.5 | 110.70 ± 0.74 |
| Longitude of asc. node[b] (deg) | $\Omega$ | 109.5 ± 1.1 | 71.2 ± 1.1 |
| Longitude of periapsis[b] (deg) | $\varpi$ | 151.70 ± 0.97 | 116.9 ± 1.5 |
| Derived parameters: | | | |
| Standard gravitational parameter $GM_{sys}$ (km$^3$ day$^{-2}$) | $\mu$ | 0.2063 ± 0.0021 | 0.2092 ± 0.0020 |
| System mass (10$^{18}$ kg) | $M_{sys}$ | 3.090 ± 0.031 | 3.135 ± 0.030 |
| Orbit pole right ascension[b] (deg) | $\alpha_{pole}$ | 19.5 ± 1.0 | 341.2 ± 1.1 |
| Orbit pole declination[b] (deg) | $\delta_{pole}$ | −37.78 ± 0.61 | +44.38 ± 0.67 |
| Orbit pole ecliptic longitude (deg) | $\lambda_{pole}$ | 359.9 ± 1.1 | 5.7 ± 1.2 |
| Orbit pole ecliptic latitude (deg) | $\beta_{pole}$ | −41.83 ± 0.58 | +47.16 ± 0.57 |
| Next mutual event season | | 2083 | 2084 |

Table notes:

[a.] Elements are for secondary relative to primary. The average sky plane residual for Orbit 1 is 0.36 mas and the maximum is 0.90 mas.

[b.] Referenced to J2000 equatorial frame.

[c.] The epoch is Julian date 2454200.0 (2007 April 9 12:00 UT).

**Table 4.** Orbital parameters and 1-$\sigma$ uncertainties for 2003 TJ$_{58}$.

| Parameter | | Orbit 1 ($\chi^2 = 1.9$) | Orbit 2 ($\chi^2 = 6.4$) |
|---|---|---|---|
| Fitted elements:[a] | | | |
| Period (days) | $P$ | 137.32 ± 0.19 | 137.32 ± 0.19 |
| Semimajor axis (km) | $a$ | 3799 ± 54 | 3728 ± 44 |
| Eccentricity | $e$ | 0.528 ± 0.011 | 0.529 ± 0.011 |
| Inclination[b] (deg) | $i$ | 38.1 ± 2.1 | 96.1 ± 2.0 |
| Mean longitude[b] at epoch[c] (deg) | $\epsilon$ | 56.0 ± 2.2 | 31.9 ± 1.9 |
| Longitude of asc. node[b] (deg) | $\Omega$ | 194.6 ± 4.2 | 150.8 ± 2.8 |
| Longitude of periapsis[b] (deg) | $\varpi$ | 84.56 ± 0.73 | 61.9 ± 3.6 |
| Derived parameters: | | | |
| Standard gravitational parameter $GM_{sys}$ (km$^3$ day$^{-2}$) | $\mu$ | 0.01537 ± 0.00066 | 0.01453 ± 0.00052 |
| System mass (10$^{17}$ kg) | $M_{sys}$ | 2.304 ± 0.099 | 2.178 ± 0.078 |
| Orbit pole right ascension[b] (deg) | $\alpha_{pole}$ | 104.6 ± 4.3 | 60.8 ± 2.9 |
| Orbit pole declination[b] (deg) | $\delta_{pole}$ | +51.9 ± 2.2 | −6.1 ± 2.1 |
| Orbit pole ecliptic longitude (deg) | $\lambda_{pole}$ | 100.2 ± 3.1 | 57.2 ± 3.0 |

| Parameter | | Orbit 1 ($\chi^2 = 1.9$) | Orbit 2 ($\chi^2 = 6.4$) |
|---|---|---|---|
| Orbit pole ecliptic latitude (deg) | $\beta_{pole}$ | $+29.0 \pm 2.1$ | $-26.3 \pm 2.0$ |
| Next mutual event season | | 2098 | 2059 |

Table notes:

a. Elements are for secondary relative to primary. The average sky plane residual for Orbit 1 is 1.1 mas and the maximum is 2.2 mas.

b. Referenced to J2000 equatorial frame.

c. The epoch is Julian date 2454300.0 (2007 July 18 12:00 UT).

**Table 5.** Orbital parameters and 1-$\sigma$ uncertainties for 2001 XR$_{254}$.

| Parameter | | Orbit 1 ($\chi^2 = 6.2$) | Orbit 2 ($\chi^2 = 14$) |
|---|---|---|---|
| Fitted elements:[a] | | | |
| Period (days) | $P$ | $125.61 \pm 0.12$ | $125.61 \pm 0.13$ |
| Semimajor axis (km) | $a$ | $9326 \pm 75$ | $9211 \pm 69$ |
| Eccentricity | $e$ | $0.5559 \pm 0.0049$ | $0.5474 \pm 0.0049$ |
| Inclination[b] (deg) | $i$ | $41.07 \pm 0.22$ | $154.50 \pm 0.22$ |
| Mean longitude[b] at epoch[c] (deg) | $\epsilon$ | $151.90 \pm 0.38$ | $11.2 \pm 1.1$ |
| Longitude of asc. node[b] (deg) | $\Omega$ | $341.16 \pm 0.33$ | $125.18 \pm 0.55$ |
| Longitude of periapsis[b] (deg) | $\varpi$ | $246.4 \pm 1.0$ | $103.3 \pm 2.0$ |
| Derived parameters: | | | |
| Standard gravitational parameter $GM_{sys}$ (km$^3$ day$^{-2}$) | $\mu$ | $0.2718 \pm 0.0065$ | $0.2619 \pm 0.0059$ |
| System mass ($10^{18}$ kg) | $M_{sys}$ | $4.073 \pm 0.098$ | $3.924 \pm 0.089$ |
| Orbit pole right ascension[b] (deg) | $\alpha_{pole}$ | $251.16 \pm 0.34$ | $35.18 \pm 0.55$ |
| Orbit pole declination[b] (deg) | $\delta_{pole}$ | $+48.93 \pm 0.22$ | $-64.50 \pm 0.21$ |
| Orbit pole ecliptic longitude (deg) | $\lambda_{pole}$ | $231.91 \pm 0.66$ | $339.51 \pm 0.55$ |
| Orbit pole ecliptic latitude (deg) | $\beta_{pole}$ | $+69.89 \pm 0.22$ | $-67.93 \pm 0.24$ |
| Next mutual event season | | 2036 | 2120 |

Table notes:

a. Elements are for secondary relative to primary. Excluding the final observation, the average sky plane residual for Orbit 1 is 1.0 mas and the maximum is 2.2 mas. The final observation (made when the two bodies were poorly resolved) has a much larger residual of 9.8 mas. With $\chi^2 = 13.6$, Orbit 2 is only excluded at 1.9-$\sigma$ confidence, below our 3-$\sigma$ threshold.

b. Referenced to J2000 equatorial frame.

c. The epoch is Julian date 2454300.0 (2007 July 18 12:00 UT).

**Table 6.** Orbital parameters and 1-$\sigma$ uncertainties for 1999 OJ$_4$.

| Parameter | | Orbit 1 ($\chi^2 = 8.6$) | Orbit 2 ($\chi^2 = 16$) |
|---|---|---|---|
| Fitted elements:[a] | | | |
| Period (days) | $P$ | 84.093 ± 0.016 | 84.136 ± 0.016 |
| Semimajor axis (km) | $a$ | 3303 ± 24 | 3225 ± 18 |
| Eccentricity | $e$ | 0.3691 ± 0.0083 | 0.3602 ± 0.0078 |
| Inclination[b] (deg) | $i$ | 53.8 ± 1.2 | 99.8 ± 1.5 |
| Mean longitude[b] at epoch[c] (deg) | $\epsilon$ | 37.6 ± 1.1 | 349.0 ± 1.9 |
| Longitude of asc. node[b] (deg) | $\Omega$ | 275.8 ± 2.2 | 210.2 ± 1.6 |
| Longitude of periapsis[b] (deg) | $\varpi$ | 329.76 ± 0.67 | 281.9 ± 2.5 |
| Derived parameters: | | | |
| Standard gravitational parameter $GM_{sys}$ (km$^3$ day$^{-2}$) | $\mu$ | 0.02696 ± 0.00058 | 0.02507 ± 0.00043 |
| System mass (10$^{17}$ kg) | $M_{sys}$ | 4.039 ± 0.087 | 3.756 ± 0.065 |
| Orbit pole right ascension[b] (deg) | $\alpha_{pole}$ | 185.8 ± 2.2 | 120.2 ± 1.6 |
| Orbit pole declination[b] (deg) | $\delta_{pole}$ | +36.2 ± 1.2 | −9.8 ± 1.6 |
| Orbit pole ecliptic longitude (deg) | $\lambda_{pole}$ | 168.7 ± 1.6 | 124.9 ± 1.6 |
| Orbit pole ecliptic latitude (deg) | $\beta_{pole}$ | +35.1 ± 1.7 | −29.6 ± 1.8 |
| Next mutual event season | | 2078 | 2048 |

Table notes:

[a.] Elements are for secondary relative to primary. Excluding the lower resolution NICMOS observation, the average sky plane residual for Orbit 1 is 1.4 mas and the maximum is 3.7 mas. The Orbit 1 residual for the NICMOS observation is 25 mas. With $\chi^2 = 16.3$, Orbit 2 is only excluded at 2.3-$\sigma$ confidence, below our 3-$\sigma$ threshold.

[b.] Referenced to J2000 equatorial frame.

[c.] The epoch is Julian date 2454000.0 (2006 September 21 12:00 UT).

**Table 7.** Orbital parameters and 1-$\sigma$ uncertainties for (134860) 2000 OJ$_{67}$.

| Parameter | | Orbit 1 ($\chi^2 = 2.3$) | Orbit 2 ($\chi^2 = 2.5$) |
|---|---|---|---|
| Fitted elements:[a] | | | |
| Period (days) | $P$ | 22.0412 ± 0.0040 | 22.0412 ± 0.0036 |
| Semimajor axis (km) | $a$ | 2361 ± 36 | 2352 ± 35 |
| Eccentricity | $e$ | 0.090 ± 0.022 | 0.085 ± 0.021 |
| Inclination[b] (deg) | $i$ | 84.6 ± 3.0 | 73.8 ± 2.9 |
| Mean longitude[b] at epoch[c] (deg) | $\epsilon$ | 71.0 ± 3.2 | 22.4 ± 3.0 |
| Longitude of asc. node[b] (deg) | $\Omega$ | 272.9 ± 3.1 | 212.2 ± 3.3 |
| Longitude of periapsis[b] (deg) | $\varpi$ | 39 ± 14 | 349 ± 17 |

| Parameter | | Orbit 1 ($\chi^2 = 2.3$) | Orbit 2 ($\chi^2 = 2.5$) |
|---|---|---|---|
| Derived parameters: | | | |
| Standard gravitational parameter $GM_{sys}$ (km$^3$ day$^{-2}$) | $\mu$ | $0.1433 \pm 0.0066$ | $0.1417 \pm 0.0063$ |
| System mass ($10^{18}$ kg) | $M_{sys}$ | $2.146 \pm 0.099$ | $2.123 \pm 0.094$ |
| Orbit pole right ascension[b] (deg) | $\alpha_{pole}$ | $182.9 \pm 3.1$ | $122.2 \pm 3.2$ |
| Orbit pole declination[b] (deg) | $\delta_{pole}$ | $+5.4 \pm 3.1$ | $+16.2 \pm 3.0$ |
| Orbit pole ecliptic longitude (deg) | $\lambda_{pole}$ | $180.5 \pm 3.0$ | $120.9 \pm 2.9$ |
| Orbit pole ecliptic latitude (deg) | $\beta_{pole}$ | $+6.1 \pm 3.2$ | $-3.8 \pm 3.2$ |
| Next mutual event season | | 2104 | 2056 |

Table notes:

[a] Elements are for secondary relative to primary. The average sky plane residual for Orbit 1 is 1.3 mas and the maximum is 1.9 mas.

[b] Referenced to J2000 equatorial frame.

[c] The epoch is Julian date 2454000.0 (2006 September 21 12:00 UT).

**Table 8.** Orbital parameters and 1-$\sigma$ uncertainties for 2004 PB$_{108}$.

| Parameter | | Orbit 1 ($\chi^2 = 2.2$) | Orbit 2 ($\chi^2 = 6.6$) |
|---|---|---|---|
| Fitted elements:[a] | | | |
| Period (days) | $P$ | $97.017 \pm 0.070$ | $97.077 \pm 0.069$ |
| Semimajor axis (km) | $a$ | $10400 \pm 130$ | $10550 \pm 130$ |
| Eccentricity | $e$ | $0.4372 \pm 0.0074$ | $0.4455 \pm 0.0074$ |
| Inclination[b] (deg) | $i$ | $89.0 \pm 1.1$ | $106.55 \pm 0.99$ |
| Mean longitude[b] at epoch[c] (deg) | $\epsilon$ | $168.49 \pm 0.63$ | $59.2 \pm 1.5$ |
| Longitude of asc. node[b] (deg) | $\Omega$ | $121.99 \pm 0.75$ | $30.19 \pm 0.86$ |
| Longitude of periapsis[b] (deg) | $\varpi$ | $351.92 \pm 0.53$ | $242.1 \pm 1.6$ |
| Derived parameters: | | | |
| Standard gravitational parameter $GM_{sys}$ (km$^3$ day$^{-2}$) | $\mu$ | $0.632 \pm 0.023$ | $0.660 \pm 0.025$ |
| System mass ($10^{18}$ kg) | $M_{sys}$ | $9.47 \pm 0.35$ | $9.88 \pm 0.37$ |
| Orbit pole right ascension[b] (deg) | $\alpha_{pole}$ | $31.99 \pm 0.75$ | $300.19 \pm 0.85$ |
| Orbit pole declination[b] (deg) | $\delta_{pole}$ | $+1.0 \pm 1.1$ | $-16.55 \pm 0.98$ |
| Orbit pole ecliptic longitude (deg) | $\lambda_{pole}$ | $30.16 \pm 0.89$ | $298.90 \pm 0.90$ |
| Orbit pole ecliptic latitude (deg) | $\beta_{pole}$ | $-11.2 \pm 1.0$ | $+3.91 \pm 0.91$ |
| Next mutual event season | | 2103 | 2039 |

Table notes:

<sup>a.</sup> Elements are for secondary relative to primary. The average sky plane residual for Orbit 1 is 1.6 mas and the maximum is 3.4 mas.

<sup>b.</sup> Referenced to J2000 equatorial frame.

<sup>c.</sup> The epoch is Julian date 2454200.0 (2007 April 9 12:00 UT).

The $\chi^2$ values shown in Tables 3-8 are computed as $\chi^2 = \sum((\text{model} - \text{data})/\sigma)^2$, where $\sigma$ are the astrometric uncertainties. The reduced $\chi^2$, usually denoted $\chi_\nu^2$, can be computed by dividing $\chi^2$ by $\nu = 2n - 7$, where $n$ is the number of astrometric observations, each of which provides two constraints (see Table 2), and seven is the number of fitted parameters. An important caveat is that separate observational visits do not necessarily provide strictly independent constraints from one another, meaning that $\nu$ could actually be somewhat less than $2n - 7$ and $\chi^2$ values can be smaller than the expected values, especially for systems having fewer observations. In these cases, we rely on the Grundy et al. (2008) Monte Carlo method described earlier to rule out the existence of other orbital solutions.

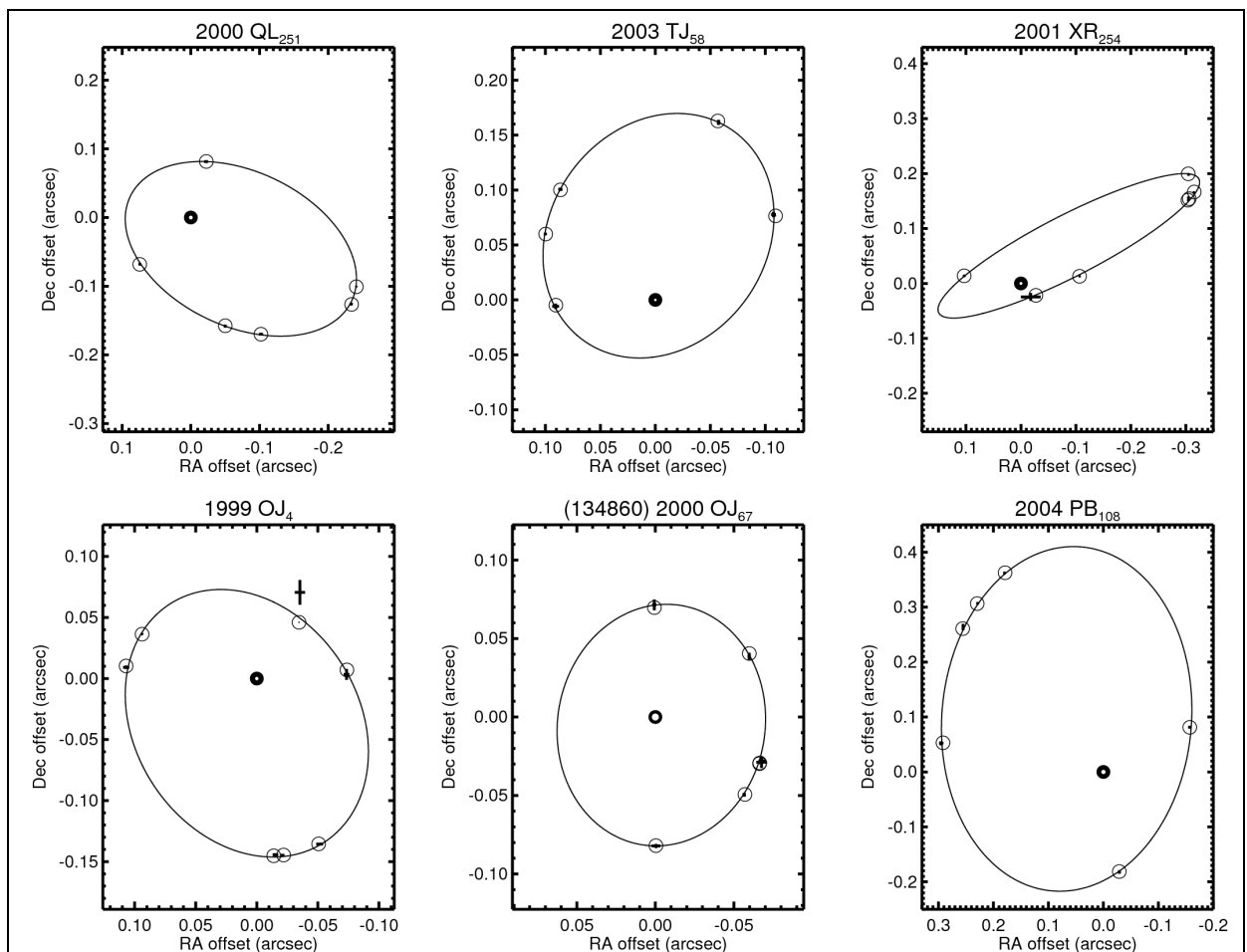

**Fig. 3:** Relative astrometry and orbit fits projected onto the sky plane. Circles at (0,0) represent the primary, with white centers scaled to the primary's estimated size. Small points with error bars, sometimes difficult to discern, show observed relative astrometry. Open circles with central dots indicate the best fit orbit solution's predicted positions at the observation times. Large ovals represent the sky plane projection of the best fit orbit at the mean time of the observations. Parallaxes from Earth and object motion around the Sun change the orbit's sky plane projection over time, causing the observed and model points to deviate slightly from these instantaneous projections, but not yet enough to distinguish between the two mirror solutions.

As mentioned previously, we report two related orbit solutions for each TNB, mirror images of one another through the sky plane at the time of observations. To break the ambiguity between these two solutions will require one or more additional observations at a later date to take advantage of the slowly changing view of the system as the TNB and Earth orbit the Sun. With the astrometric precision of HST, a few years delay is sometimes sufficient to exclude one of the two orbit solutions, but this depends strongly on the orientation, shape, and size of the orbit as well as the specific orbital longitudes sampled by the observations (see Grundy et al. 2008). When an additional observation can raise the $\chi^2$ statistic for the wrong orbit above the 3-$\sigma$ confidence threshold for exclusion (which can be calculated using the incomplete gamma function; Press et al. 1992), that orbit can be formally excluded. At present, none of the mirror solutions for the six TNBs in this paper reach that threshold. The closest is 1999 OJ$_4$, for which Orbit 2 can be excluded at 2.3-$\sigma$ confidence. In the mean time, for each TNB system we adopt $P$, $a$, and $e$ values intermediate between the two mirror solutions, with uncertainties inflated to encompass the full range of 1-$\sigma$ uncertainties for the two solutions. These adopted values appear in Table 9, along with mean $V$-band photometric properties.

**Table 9.** Adopted orbital elements and photometric properties.

| TNB system | Period $P$ (days)[a] | Semimajor axis $a$ (km)[a] | Eccentricity $e$[a] | Combined $H_v$[b] (mag) | $\Delta_{mag}$[b] (mag) |
|---|---|---|---|---|---|
| 2000 QL$_{251}$ | 56.451 ± 0.025 | 5002 ± 27 | 0.4871 ± 0.0065 | 7.148 ± 0.016 | 0.073 ± 0.048 |
| 2003 TJ$_{58}$ | 137.32 ± 0.19 | 3768 ± 85 | 0.529 ± 0.011 | 7.448 ± 0.024 | 0.519 ± 0.031 |
| 2001 XR$_{254}$ | 125.61 ± 0.13 | 9271 ± 130 | 0.5516 ± 0.0091 | 6.030 ± 0.017 | 0.415 ± 0.064 |
| 1999 OJ$_4$ | 84.115 ± 0.038 | 3267 ± 60 | 0.365 ± 0.012 | 7.344 ± 0.012 | 0.088 ± 0.020 |
| (134860) 2000 OJ$_{67}$ | 22.0412 ± 0.0040 | 2357 ± 40 | 0.088 ± 0.025 | 6.471 ± 0.010 | 0.530 ± 0.073 |
| 2004 PB$_{108}$ | 97.05 ± 0.10 | 10480 ± 200 | 0.441 ± 0.012 | 7.083 ± 0.016 | 1.321 ± 0.032 |

Table notes:

[a.] Orbital element values for the orbit of the secondary about the primary along with uncertainties here are chosen such that symmetric error bars encompass the 1-$\sigma$ uncertainties of the two mirror orbit solutions for each TNB.

[b.] Average absolute $V$ magnitude $H_V$ of the combined light from both bodies and average $V$ magnitude difference between primary and secondary $\Delta_{mag}$ computed from the ensemble of WFPC2 *F606W* observations as described by a companion paper (Benecchi et al. 2009) assuming $G = 0.15$ in the $H$ and $G$ magnitude system (Bowell et al. 1989). These numbers differ somewhat from numbers in Benecchi et al. (2009) Table 3 because those numbers were based only on the sub-set of visits consisting of interleaved observations in *F606W* and *F814W* filters (typically just one visit per system in Cycle 16 program 11178).

## 3. Results

System masses for our TNBs are computed according to

$$M_{sys} = \frac{4\pi^2 a^3}{G P^2}, \qquad (1)$$

where $G$ is the gravitational constant (we use the CODATA 2006 value $G = 6.6742 \times 10^{-11}$ m$^3$ s$^{-2}$

kg$^{-1}$). The semimajor axis $a$ is usually the dominant source of uncertainty in $M_{sys}$ because it tends to be constrained to fewer significant digits by our observations than the period $P$ and because it is raised to a higher power.

**Table 10.** Derived quantities.

| Object designation | System mass[a] $M_{sys}$ (kg) | Primary Radius (km)[b] | $V$ geometric albedo $A_p$[b] | Orbital angular momentum $J_{orb}/J'$ | Hill radius $r_H$ (km) | $a/r_H$ |
|---|---|---|---|---|---|---|
| 2000 QL$_{251}$ | $(3.112 \pm 0.051) \times 10^{18}$ | 57 - 91 | 0.04 - 0.10 | 1.7 - 2.1 | 460,000 | 0.011 |
| 2003 TJ$_{58}$ | $(2.25 \pm 0.15) \times 10^{17}$ | 25 - 40 | 0.16 - 0.41 | 2.2 - 2.7 | 200,000 | 0.019 |
| 2001 XR$_{254}$ | $(4.00 \pm 0.17) \times 10^{18}$ | 65 - 104 | 0.09 - 0.22 | 2.1 - 2.7 | 550,000 | 0.017 |
| 1999 OJ$_4$ | $(3.91 \pm 0.22) \times 10^{17}$ | 29 - 46 | 0.12 - 0.31 | 1.9 - 2.4 | 230,000 | 0.014 |
| 2000 OJ$_{67}$ | $(2.14 \pm 0.11) \times 10^{18}$ | 53 - 85 | 0.09 - 0.23 | 1.2 - 1.5 | 450,000 | 0.005 |
| 2004 PB$_{108}$ | $(9.68 \pm 0.57) \times 10^{18}$ | 93 - 148 | 0.02 - 0.05 | 1.6 - 2.0 | 710,000 | 0.015 |

Table notes:

[a.] System masses $M_{sys}$ are based on the CODATA 2006 value of the gravitational constant $G = 6.6742 \times 10^{-11}$ m$^3$ s$^{-2}$ kg$^{-1}$.

[b.] Radii and albedos are based on $M_{sys}$ and a plausible density range of 0.5 to 2.0 g cm$^{-3}$, assuming both bodies share the same albedo and density. Higher densities would permit smaller radii and higher albedos, while lower densities would allow larger radii and lower albedos.

The masses and photometric properties in Table 9 enable us to estimate a number of additional properties of these systems, as shown in Table 10. For an assumed range of plausible densities, we can constrain the sizes and albedos of the bodies to lie within a corresponding range. The plausible density range is taken to be 0.5 to 2.0 g cm$^{-3}$ (e.g., Lacerda and Jewitt 2007; Grundy et al. 2008). We assume the two components share the same density and albedo, with the volume implied by the known mass and assumed density being shared between them such that their projected area ratio matches their average flux ratio and their total projected area gives the observed average visual magnitude of the combined system, as described by Grundy et al. (2005). Photometric values used in this calculation were derived according to techniques described in a companion paper (Benecchi et al. 2009). Four of the six TNBs belong to the dynamically "Cold" Classical disk (see Table 1). These four all have minimum albedos of 0.09 or higher, consistent with the recent finding that Cold Classical objects are characterized by elevated albedos (Brucker et al. 2009) in addition to their red colors and high rates of binarity (e.g., Doressoundiram et al. 2008; Noll et al. 2008b). The other two objects, both having more excited heliocentric orbits, are found to have lower albedos.

From the system mass and orbital parameters, we can estimate the orbital angular momentum $J_{orb}$. If we knew the densities and spin periods of the individual bodies, we could also compute the spin component of the angular momentum $J_{spin}$. The total angular momentum $J = J_{orb} + J_{spin}$ is conventionally normalized to $J' = \sqrt{G M_{sys}^3 R_{eff}}$ where $R_{eff}$ is the radius of a single spherical body with the same total mass (e.g., Noll et al. 2008a). This specific angular momentum $J/J'$ has been used to constrain the origins of binary systems such as the Pluto-Charon system (e.g., Canup 2005; Chiang et al. 2007). Binaries produced via impact disruption should have $J/J' < 0.8$, at least if impact speeds do not greatly exceed escape velocities. Not knowing the spin periods of the individual components, we choose to omit $J_{spin}$, and tabulate $J_{orb}/J'$ in Table 10. As before, the plausible density range range results in a plausible range of $J_{orb}/J'$ values. The orbital angular momentum suggests that these six TNB systems would not have formed by impact disruption. However, these bodies have escape velocities on the order of tens of meters per second. Impactors arrive at much higher speeds, at least in the present-day transneptunian collisional environment (e.g., Stern and Kenyon 2003). The $J/J' < 0.8$ criterion may need to be modified for this regime.

Hill radii $r_H$ can be computed from $M_{sys}$ in conjunction with the solar mass $M_\odot$, the heliocentric semimajor axis $a_\odot$, and heliocentric eccentricity $e_\odot$, according to

$$r_H = a_\odot (1-e_\odot) \left( \frac{M_{sys}}{3 M_\odot} \right)^{\frac{1}{3}}. \qquad (2)$$

The mutual orbits of the six TNBs have semimajor axes between 0.5 and 2 percent of their Hill radii, as measured by $a/r_H$ in Table 10. Orbits this tight are very unlikely to be disrupted by the gravitational influence of distant perturbers. However, they could be modified by other TNOs happening to pass nearby, as discussed by Petit and Mousis (2004).

Five of the six systems have similar eccentricities, between 0.37 and 0.55. This observation may offer another useful constraint on possible formation scenarios, but we are not aware of any published binary formation mechanisms which produce eccentricities clustered in this range. The one system with a much lower eccentricity is (134860) 2000 OJ$_{67}$. Of the six, this system also has the shortest period (22 days) and the smallest semimajor axis (2360 km). Tidal interactions depend strongly on the proximity of the two bodies relative to their sizes, suggesting the possibility of tidal evolution circularizing an initially more eccentric orbit, as was proposed for the (65489) Ceto-Phorcys system (Grundy et al. 2007; see also Noll et al. 2008a and references therein). While the distribution of eccentricities in this sample is intriguing, the sample is likely to be biased in various ways and may not be representative of the distribution of eccentricities among TNB systems in general.

Although the orientations of all six orbits remain ambiguous as to which of the two mirror solutions is correct, most are highly inclined with respect to the ecliptic plane, as indicated by the low ecliptic latitudes of their poles $\beta_{pole}$ in Tables 3-8. In particular, the mutual orbits of (134860) 2000 OJ$_{67}$ and of 2004 PB$_{108}$ are nearly perpendicular to the ecliptic plane. None happens to be oriented so as to offer the prospect of mutual events within the next two decades.

## 4. Conclusion

We used the Planetary Camera of the Wide Field and Planetary Camera 2 (WFPC2) aboard Hubble Space Telescope to repeatedly image six known transneptunian binaries, 2000 QL$_{251}$, 2003 TJ$_{58}$, 2001 XR$_{254}$, 1999 OJ$_4$, (134860) 2000 OJ$_{67}$, and 2004 PB$_{108}$. Measurements of the time-dependent astrometry of the secondary body relative to the primary body were used to derive their mutual orbits. We found orbital periods ranging from 22 to 137 days and semimajor axes ranging from 2360 to 10500 km for these six binary systems. Orbital eccentricities of five systems were found to be between 0.37 and 0.55. (134860) 2000 OJ$_{67}$ has a lower eccentricity of 0.09. The mutual orbits enabled us to estimate the masses of the systems, which were found to range from $2.1 \times 10^{17}$ to $9.7 \times 10^{18}$ kg. These masses were combined with visible photometry to constrain the sizes and albedos of the bodies. The four dynamically cold Classical objects in our sample all have visual geometric albedos $\geq 0.09$ if their bulk densities are $\geq 0.5$ g cm$^{-3}$. The two objects in more excited heliocentric orbits have lower albedos. The angular momentum of the binary orbits is sufficiently high to favor formation via capture rather than impact disruption.

**Acknowledgments**

This work is based on NASA/ESA Hubble Space Telescope Cycle 16 program 11178 observations, using additional data from programs 9386, 10514, and 10800. Support for all of these programs was provided by NASA through grants from the Space Telescope Science Institute (STScI), which is operated by the Association of Universities for Research in Astronomy, Inc., under NASA contract NAS 5-26555. We are especially grateful to Tony Roman at STScI for his quick action in scheduling HST follow-up observations in program 11178. We thank two anonymous reviewers for constructive suggestions leading to improvements of this paper. Finally, we thank the free and open source software communities for empowering us with the software tools used to complete this project, notably Linux, the GNU tools, OpenOffice.org, MySQL, STSDAS, Evolution, GDL, Python, and FVWM.